\documentclass[aps,pra,final,10pt,twocolumn,superscriptaddress]{revtex4-2}

\usepackage{amsmath,amssymb,bm}
\usepackage{graphicx,color}
\usepackage[squaren]{SIunits}
\usepackage[english]{babel}
\usepackage{amstext}
\usepackage{amsthm}
\usepackage{latexsym}
\usepackage{array}
\usepackage{color}
\usepackage{float}
\usepackage{mhchem}
\usepackage{microtype}
 
\usepackage{multirow}
\usepackage{dcolumn}
\definecolor{url}{RGB}{0,20,160}
\usepackage[colorlinks=true,linkcolor=blue,citecolor=blue,urlcolor=url]{hyperref}
\usepackage[usenames,dvipsnames,svgnames,table]{xcolor}
\usepackage{graphicx, type1cm, lettrine}
\usepackage[utf8]{inputenc}

\newcommand{\webposabr}{\href{https://github.com/zjwang11/UnconvMat/blob/master/src_pos2aBR.tar.gz}{\texttt{POS2ABR }}}

\usepackage{natbib}
\makeatletter
\def\frutiger{cmss10 }
\def\frutigerbold{cmssbx10 }
=\frutigerbold at 14pt
=\frutiger at 8pt
=\frutigerbold at 12pt
=\frutigerbold at10pt
=\frutigerbold at 8pt
=\frutiger at 8pt
=\frutigerbold at 31pt

\def\@caption@tabnum@sep{\figtextfont{{ }{\bf\textbar}{ }}}%
\def\fnum@table{{\bf\tablename~\thetable}}

\def\@caption@fignum@sep{\figtextfont{{ }{\bf\textbar}{ }}}%
\def\fnum@figure{{\bf\figurename~\thefigure}}
\renewenvironment{figure}{\@float{figure}\def\textbf##1{{\fignumfont ##1}}\def\bf{\fignumfont}}{\end@float}

\def\@startsection#1#2#3#4#5#6{%
	\if@noskipsec\leavevmode\fi
	\par\@tempskipa #4\relax
	\@afterindenttrue
	\ifdim\@tempskipa <\z@
	\@tempskipa -\@tempskipa \@afterindentfalse
	\fi\if@nobreak\everypar{}%
	\else\addpenalty\@secpenalty\addvspace\@tempskipa\fi
	\@ifstar{\@ssect{#3}{#4}{#5}{#6}}{\@dblarg{\@sect{#1}{#2}{#3}{#4}{#5}{#6}}}}

\def\@sect#1#2#3#4#5#6[#7]#8{%
	\ifnum #2>0
	\let\@svsec\@empty
	\else\refstepcounter{#1}\protected@edef\@svsec{\@seccntformat{#1}\relax}\fi
	\@tempskipa #5\relax
	\ifdim\@tempskipa>\z@
	\begingroup#6{\@hangfrom{\hskip #3\relax\@svsec}%
		\interlinepenalty \@M #8\@@par}\endgroup
	\csname #1mark\endcsname{#7}%
	\addcontentsline{toc}{#1}{%
		\ifnum #2>\c@secnumdepth\else
		\protect\numberline{\csname the#1\endcsname}\fi #7}%
	\else\def\@svsechd{#6{\hskip #3\relax
			\@svsec #8\ifnum#2=2.\fi}%
		\csname #1mark\endcsname{#7}%
		\addcontentsline{toc}{#1}{%
			\ifnum #2>\c@secnumdepth \else
			\protect\numberline{\csname the#1\endcsname}\fi #7}}%
	\fi\@xsect{#5}}

\renewcommand\section{\@startsection {section}{1}{\z@}%
	{-10pt \@plus -1ex \@minus -.2ex}{.5ex }{\normalfont\Large\bfseries\sectionfont}}
\renewcommand\subsection{\@startsection{subsection}{2}{\z@}%
	{10pt\@plus 1ex \@minus .2ex}{-0.5ex \@plus .2ex}{\normalfont\large\bfseries\subsectionfont}}

\def\frontmatter@title@format{\titlefont\centering}%
\def\frontmatter@title@below{\addvspace{-5pt}}%

\renewcommand\NAT@biblabelnum[1]{#1.}
\renewcommand\NAT@citesuper[3]{\ifNAT@swa
	\unskip\hspace{1\p@}\textsuperscript{(#1)}%
	\if\relax#3\relax\else\ (#3)\fi\else (#1)\fi\endgroup}

\newcommand*\bib@heading{%
	\section{\refname}
	\fontsize{8}{10}\selectfont
}
\newcommand*\@openbib@code{%
	\advance\leftmargin\bibindent
	\itemindent -\bibindent
	\listparindent \itemindent
	\parsep \z@
}%
\newdimen\bibindent
\usepackage[usenames,dvipsnames,svgnames,table]{xcolor}
\definecolor{col1}{rgb}{0.0, 0.30, 1.0}
\definecolor{col2}{rgb}{0.9, 0.0, 0.30}

\makeatletter

\begin{document}

\title{Forbidden p-d Orbital Coupling Accelerates High-Power-Factor Materials Discovery}

\author{Wu Xiong}
\affiliation{Key Laboratory of Advanced Materials and Devices for Post-Moore Chips, Ministry of Education, University of Science and Technology Beijing, Beijing 100083, China}
\affiliation{School of Mathematics and Physics, University of Science and Technology Beijing, Beijing 100083, China}

\author{Zhongjuan Han}
\affiliation{Key Laboratory of Advanced Materials and Devices for Post-Moore Chips, Ministry of Education, University of Science and Technology Beijing, Beijing 100083, China}
\affiliation{School of Mathematics and Physics, University of Science and Technology Beijing, Beijing 100083, China}

\author{Zhonghao Xia}
\affiliation{Key Laboratory of Advanced Materials and Devices for Post-Moore Chips, Ministry of Education, University of Science and Technology Beijing, Beijing 100083, China}
\affiliation{School of Mathematics and Physics, University of Science and Technology Beijing, Beijing 100083, China}

\author{Zhilong Yang}
\email{zhlyang@ustb.edu.cn}
\affiliation{Key Laboratory of Advanced Materials and Devices for Post-Moore Chips, Ministry of Education, University of Science and Technology Beijing, Beijing 100083, China}
\affiliation{School of Mathematics and Physics, University of Science and Technology Beijing, Beijing 100083, China}

\author{Jiangang He}
\email{jghe2021@ustb.edu.cn}
\affiliation{Key Laboratory of Advanced Materials and Devices for Post-Moore Chips, Ministry of Education, University of Science and Technology Beijing, Beijing 100083, China}
\affiliation{School of Mathematics and Physics, University of Science and Technology Beijing, Beijing 100083, China}

\date{\today}

	\begin{abstract}
		The intrinsic entanglement between electrical conductivity ($\sigma$) and the Seebeck coefficient ($S$) significantly constrains power factor (PF) enhancement in thermoelectric (TE) materials. While high valley degeneracy ($N_{\mathrm{vk}}$) effectively balances $\sigma$ and $S$ to improve PF, identifying compounds with high $N_{\mathrm{vk}}$ remains challenging. In this study, we develop an effective approach to rapid discover $p$-type semiconductors with high $N_{\mathrm{vk}}$ through manipulating anion-$p$ and cation-$d$ orbital coupling. By prohibiting $p$-$d$ orbital coupling at the $\Gamma$ point, the valence band maximum shifts away from the $\Gamma$ point (where $N_{\mathrm{vk}}$=1), thereby increasing $N_{\mathrm{vk}}$. Through the examination of the common irreducible representations of anion-$p$ and cation-$d$ orbitals at the $\Gamma$ point, we identify 7 compounds with $N_{\mathrm{vk}}$ $\ge$ 6 from 921 binary and ternary semiconductors. First-principles calculations with electron-phonon coupling demonstrate that PtP$_2$, PtAs$_2$, and PtS$_2$ exhibit exceptionally high PFs of 130, 127, and 82 $\mu$Wcm$^{-1}$K$^{-2}$ at 300K, respectively, which are three to five times higher than those of the well-studied TE materials. This work not only elucidates the underlying mechanism of high $N_{\mathrm{vk}}$ formation through group theory, but also establishes an efficient high-PF material discovery paradigm, extended to more complex systems.
	\end{abstract}
	
	\maketitle
	
	\noindent TE materials facilitate the direct conversion between electrical and thermal energy, presenting promising applications in the transforming of (waste) heat into electrical energy and the provision of solid-state cooling~\cite{goldsmid2010introduction,uher2016materials,science.1159725,2022Chemical,D2EE02408J,hzj}. The efficiency of energy-to-electricity conversion is intricately linked to the dimensionless figure of merit ($ZT$), which is defined as $ZT$ = $S^2$$\sigma$$T$/($\kappa_{\mathrm{L}}$ + $\kappa_{\mathrm{e}}$), where $S$, $\sigma$, $\kappa_{\mathrm{L}}$, $\kappa_{\mathrm{e}}$, and $T$ are Seebeck coefficient, electrical conductivity, lattice thermal conductivity, electron thermal conductivity, and absolute temperature, respectively. The deep entanglement among these parameters prevents simultaneous optimization, thereby limiting the energy conversion efficiency of TE materials~\cite{SHITTU2020115075,ZEIER201723,zhang2014organic}. Enhancing $ZT$ necessitates either an increase in the power factor (PF, $S^2\sigma$) or a reduction in both $\kappa_{\mathrm{L}}$ and $\kappa_{\mathrm{e}}$. Given that $\kappa_{\mathrm{e}}$ is proportional to $\sigma$, according to the Wiedemann-Franz law~\cite{solidstatephysics}, the primary strategy for improving $ZT$ involves reducing $\kappa_{\rm L}$ while simultaneously increasing the PF. In the framework of kinetic theory~\cite{2004Thermal}, $\kappa_{\rm L}$ can be expressed as $\frac{1}{3}C_v\nu_g^2 \tau$, where $C_v$, $\nu_g$, and $\tau$ are heat capacity, phonon group velocity, and phonon relaxation time, respectively. Since $\nu_g$ is proportional to $\sqrt{k/M}$, where $k$ denotes chemical bond strength and $M$ represents atomic mass, materials with short $\tau$, low $C_v$, weak $k$, and relatively large $M$ typically exhibit low $\kappa_{\rm L}$. To date, numerous strategies have been employed to suppress $\kappa_{\mathrm{L}}$, including defect engineering~\cite{Mao03042018,2018Nano,2008High}, nano-structure precipitates~\cite{doi:10.1126/science.1092963,2012High,doi:10.1126/science.1156446,2008Enhanced}, lone-pair electrons~\cite{PhysRevLett.107.235901,2013Lone}, rattling phonon modes~\cite{2015Impact,2016Ultralow,2021Physical}, and weak chemical~\cite{https://doi.org/10.1002/adfm.202108532,https://doi.org/10.1002/advs.202417292,he2019designing}.

	Another crucial factor for enhancing $ZT$ is the increase of the PF. Furthermore, the output power density of a device is directly correlated with PF, and thus high PF is essential for achieving high power density output~\cite{10.1063/1.3634018}. However, $\sigma$ and $S$ are heavily entangled---specifically, $\sigma$ is directly proportional to the carrier concentration ($n$), whereas $S$ exhibits an inverse relationship with $n$~\cite{2008Complex}. Band alignment is an effective strategy for enhancing PF of TE materials~\cite{moshwan2019realizing,tan2017improving,long2023band,pei2012thermoelectric,kim2017high}. This is attributed to the fact that $S$ is proportional to the effective mass of the density of states ($m_{\mathrm{d}}^*$), which in turn is proportional to both the band effective mass ($m_{\mathrm{b}}^{*}$) and band degeneracy $N_{\mathrm{v}}$ ($N_{\mathrm{v}}=N_{\mathrm{vk}}$$\times$$N_{\mathrm{vo}}$, where $N_{\mathrm{vo}}$ is the orbital degeneracy) within a single parabolic band~\cite{2018A}. Consequently, a high $N_{\mathrm{vk}}$ can significantly alleviate the competition between $S$ and $\sigma$, thereby enhancing the PF~\cite{komoto2013thermoelectric,10.1063/1.3040321,https://doi.org/10.1002/anie.202011765,doi:10.1021/acsaem.9b02131,https://doi.org/10.1002/aenm.201400600,D0TA02758H}. For instance, the converence of the second maximum of the valence band (located at the midpoint of the $\Sigma$ line, exhibiting a $N_{\mathrm{vk}}$ of 12) with the valence band maximum (VBM) at the L point ($N_{\mathrm{vk}}$=4) in PbTe, through appropriate alloying with PbSe, can result in a substantial increase in $ZT$ from 0.8 to 1.8~\cite{2011Convergence}. Similarly, the solid solution Mg$_2$Si$_{0.3}$Sn$_{0.7}$ enhances the PF to 48 $\mu$Wcm$^{-1}$K$^{-2}$ at 550 K, which is more than double that of traditional $n$-type materials~\cite{PhysRevLett.108.166601}. However, materials with intrinsically high $N_{\mathrm{v}}$ are relatively rare, and many TE materials require heavy doping or alloying, complicating material fabrication and potentially reduce stabilities. Therefore, explore semiconductors with intrinsically high $N_{\mathrm{vk}}$ in both VBM and conduction band minimum (CBM) is highly advantageous. The value of $N_{\mathrm{vk}}$ is determined by the positions of VBM or CBM within the first Brillouin zone (BZ). The $\Gamma$ point possesses the lowest $N_{\mathrm{vk}}$, equal to 1, while the zone boundaries, particularly at the midpoint of the lines connecting two high symmetry points, exhibit a high $N_{\mathrm{vk}}$ if the compound has a high symmetry.

	For most ionic semiconductors lacking transition metals, the valence and conduction bands predominantly arise from the anion-$p$ and cation-$s$ orbitals, respectively. Consequently, the VBM and CBM of these compounds are typically located at the $\Gamma$ point, as observed in NaCl and CaS, since the $p$ orbitals are aligned `downhill' and $s$ orbitals `uphill'~\cite{https://doi.org/10.1002/anie.198708461,https://doi.org/10.1002/anie.201508381}. However, in compounds containing cations with lone-pair electrons, the VBM may shift away from the $\Gamma$ point, resulting in a higher $N_{\mathrm{vk}}$. This shift occurs because $s$ orbitals are positioned `uphill', with notable examples including PbTe~\cite{PhysRevB.55.13605}, Bi$_2$Te$_3$~\cite{https://doi.org/10.1002/adfm.201900677}, and Li$_2$TlBi~\cite{he2019designing}. This strategy has been widely employed in the design of high-performance TE materials~\cite{https://doi.org/10.1002/adfm.202108532,https://doi.org/10.1002/anie.201508381}. Is there other strategy of designing semiconductors with high $N_{\mathrm{vk}}$?

	\begin{figure*}
		\includegraphics[width=1.0\linewidth,angle=0]{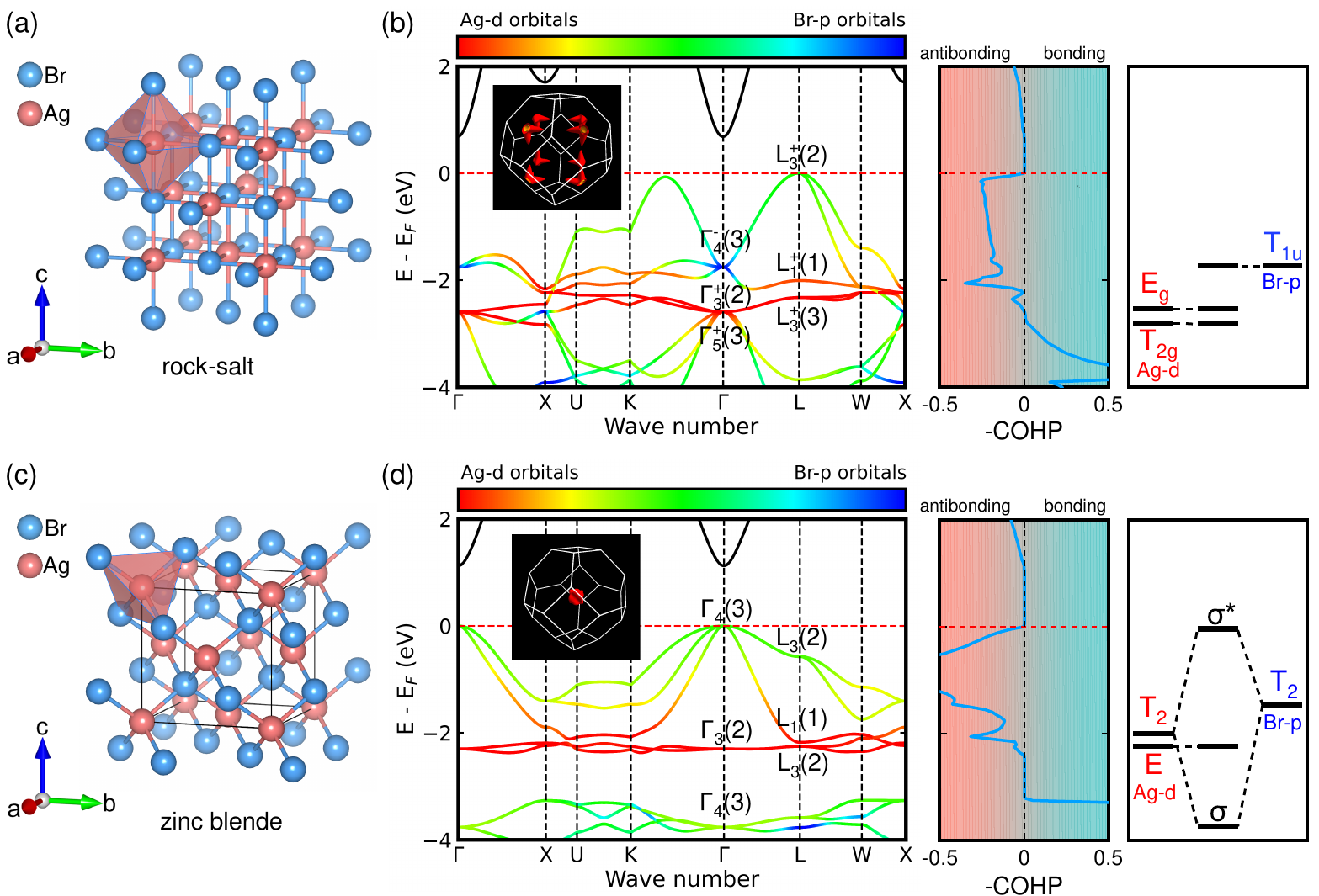}	
		\caption{(a-b) The crystal structures, orbital-electronic structures with irreps., -COHP, and MO diagrams at the $\Gamma$ point of rock-salt AgBr, respectively. (c-d) The crystal structures, orbital-projected electronic structures with irreps., -COHP, and MO diagrams at the $\Gamma$ point of zinc blende AgBr, respectively. Insets in the electronic structures are the Fermi surfaces 50 meV below the Fermi level. The color indicates the contributions from Br-$p$ and Ag-$d$ orbitals. Positive -COHP values indicate bonding interactions, while negative -COHP values indicate antibonding interactions.}
		\label{band}
	\end{figure*}

	In this work, we proposed a strategy for identifying $p$-type TE materials with high $N_{\mathrm{vk}}$ based on $p$-$d$ orbital coupling and group theory. The anti-bonding states formed through $p$-$d$ coupling located below the Fermi level can mitigate the `downhill' behavior of the anion-$p$ orbitals. If the $p$-$d$ coupling is prohibited at the $\Gamma$ point, the energy bands will `uphill' from the non-bonding states (low energy) at the $\Gamma$ point to the anti-bonding states (high energy) near the zone boundary, leading to an increased $N_{\mathrm{vk}}$. We then analyze the conditions necessary for the formation of forbidden $p$-$d$ coupling at the $\Gamma$ point by utilizing symmetry-adapted band-representation analysis. With this approach, we screen 921 binary and ternary compounds stored in the Materials Project (\href{https://legacy.materialsproject.org/}{\textcolor{blue}{www.materialproject.org}})~\cite{mp} and identify 41 compounds with high $N_{\mathrm{vk}}$. Electron transport calculations based on electron-phonon coupling demonstrate that PtP$_2$, PtAs$_2$, and PtS$_2$ exhibit maximum PF of 130, 127, and 82 $\mu$Wcm$^{-1}$K$^{-2}$ at 300K, which are 3-5 times greater than those of the well-studied TE compounds. Our results not only elucidate the underlying mechanism for the formation of high $N_{\mathrm{vk}}$ in compounds exhibiting $p$-$d$ coupling, but also facilitate the discovery of new compounds with high PF. Furthermore, this novel strategy is universal and can be extended to more complicated systems.

	\begin{table*}
		\setlength{\tabcolsep}{6pt} 
		\renewcommand{\arraystretch}{1.4} 
		\centering
		\caption{The atomic valence-electron band representations (ABRs) of rock-salt AgBr ($Fm\overline{3}m$) and zinc blende AgBr ($F\overline{4}3m$), along with the elementary band representations (EBRs) induced by them at the high-symmetry points $\Gamma$ and L. The number in parentheses denotes the dimensionality of the representation.}
		\begin{tabular}{cccccccc}
			\hline
			\hline
			& Orbitals  & WKS ($q$) & Basis                   & Irreps. ($p$) & ABRs ($p@q$) & EBRs ($\Gamma$) & EBRs (L) \\
			\hline
			& \multirow{2}{*}{Ag $d$} & \multirow{2}{*}{4$a$} & $d_{z^2}$,$d_{x^2-y^2}$ & E$_g$         & E$_g$@4$a$     & $\Gamma_3^+$(2) & L$_3^+$(2) \\
			Rock-salt AgBr &&& $d_{xy}$,$d_{xz}$,$d_{yz}$ & T$_{2g}$ & T$_{2g}$@4$a$ & $\Gamma_5^+$(3) & L$_1^+(1)$$\oplus$L$_3^+(2)$ \\	 
			& Br $p$ & 4b    & $p_x$,$p_y$,$p_z$ & T$_{1u}$ & T$_{1u}$@4b & $\Gamma_4^-$(3) & L$_1^+(1)$$\oplus$L$_3^+(2)$ \\		
			\hline
			& \multirow{2}{*}{Ag $d$} & \multirow{2}{*}{4d}   & $d_{z^2}$,$d_{x^2-y^2}$ & E & E@4d & $\Gamma_3(2)$ & L$_3(2)$ \\
			Zinc blende AgBr &&& $d_{xy}$,$d_{xz}$,$d_{yz}$ & T$_2$ & T$_2$@4d & $\Gamma_4(3)$ & L$_1(1)$$\oplus$L$_3(2)$ \\	
			& Br $p$ & 4$a$   & $p_x$,$p_y$,$p_z$ & T$_2$ & T$_2$@4$a$ & $\Gamma_4(3)$ & L$_1(1)$$\oplus$L$_3(2)$ \\
			\hline
		\end{tabular}
		\label{EBR}
	\end{table*}  
		
	\section{Results and discussion}
	\subsection{Electronic structures of rock-salt and zinc blende AgBr}
	As illustrated in Figure~\ref{band}(a), AgBr crystallizes in a rock-salt structure (space group $Fm\bar{3}m$, No. 225), in which the Ag$^{+}$ cation is octahedrally coordinated by six Br$^{-}$ anions. The band structure, depicted in Figure~\ref{band}(b), reveals that the VBM is situated at the L point of the BZ, which has $N_{\mathrm{vk}}$ of 4, while the second highest valence band is located midway between the $\Gamma$ and K points ($\Sigma$ line), which possesses $N_{\mathrm{vk}}$ of 12. In contrast, the electronic structure of zinc blende AgBr ($F\bar{4}3m$, No. 216), illustrated in Figure~\ref{band}(c), differs significantly from that of rock-salt structure. In this structure, the VBM is located at the $\Gamma$ point ($N_{\mathrm{vk}}$=1), and the highest-energy valence band exhibits a `downhill' dispersion. The high $N_{\mathrm{vk}}$ can be directly visualized by the Fermi surface; there are 8 hole pockets at the L point for rock-salt AgBr, each shared with another BZ. Conversely, in the zinc blende structure, only one pocket exists at the zone center of the first BZ, as shown in the insets of Figure~\ref{band}(b) and (d). This phenomenon can be attributed to the prohibition of $p$-$d$ orbital coupling at the $\Gamma$ point within the rock-salt structure, arising from the inverse symmetry of the octahedral crystal field, while the tetrahedral crystal field in zinc blende structure, which lacks inverse symmetry, permits $p$-$d$ coupling at the $\Gamma$ point~\cite{PhysRevB.37.8958,https://doi.org/10.1002/adfm.202108532}. However, this explanation is incomplete; other symmetry operations within the space group must also be considered.

	As shown in the orbital-projected band structures of Figure~\ref{band}(b) and (d), the valence bands of both rock-salt and zinc blende AgBr display a strong $p$-$d$ coupling, due to the high energy of the Ag-$d$ orbitals and their favorable energy match with Br-$p$ orbitals~\cite{PhysRevB.37.8958,https://doi.org/10.1002/adfm.202108532}. These valence bands are the anti-bonding states formed by the coupling of Ag-$d$ and Br-$p$ orbitals because the $d$-orbitals of Ag$^{+}$ is fully occupied. This is supported by the COHP analysis, which indicates negative -COHP value within the energy range of -2.5 eV to the Fermi level. It is important to note that the $p$-$d$ orbital coupling exhibits strongly variation along the high-symmetry points in $k$-space for both structures. In rock-salt AgBr, Ag-$d$ and Br-$p$ orbitals can't hybridize at the $\Gamma$ point, whereas strong coupling occurs at the L point, leading to an approximate 2 eV energy increase from the $\Gamma$ to the L point. In contrast, for zinc blende AgBr, $p$-$d$ orbital coupling is allowed at all $k$ points of the first BZ. Consequently, the energy of the highest valence band at the $\Gamma$ point surpasses that at the L point, indicating stronger coupling at the $\Gamma$ point. The $p$-$d$ coupling can be further confirmed by the band-decomposed charge density. As depicted in Figure~\textcolor{red}{S1}(b) and (c), both Ag and Br exhibit substantial charge density at the L point for rock-salt AgBr, while charge density at the $\Gamma$ point is only associated with Ag. In the zinc blende AgBr, although Br contributes similarly to both the $\Gamma$ and L points, Ag dispalys a slightly larger contribution to the $\Gamma$ point, as illustrated in Figure~\textcolor{red}{S1}(e) and (f).

	\begin{figure*}
		\includegraphics[width=1.0\linewidth,angle=0]{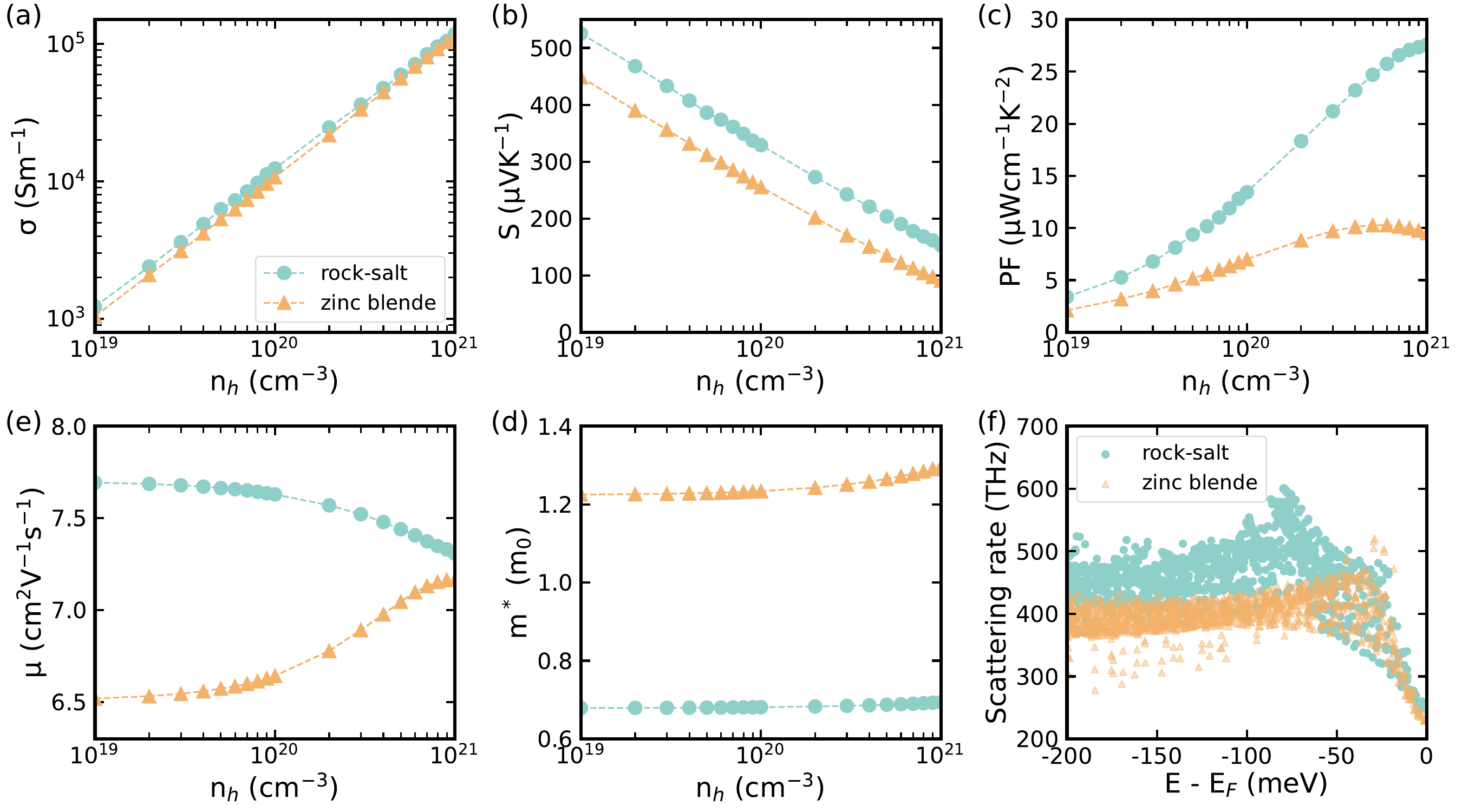}	
		\caption{(a)-(e) respectively show the variation trends of the conductivity ($\sigma$), Seebeck coefficient ($S$), power factor (PF), mobility ($\mu$), and transport effective mass ($m^*$) of rock-salt and zinc blende AgBr with $n_h$, at 300K. (f) shows the variation trend of the electron-phonon scattering rate (1/$\tau$) of these two phases with energy, at 300K.}
		\label{AgBr}
	\end{figure*}

	Under the octahedral crystal field of rock-salt AgBr, the $d$ orbitals of Ag$^+$ undergo splitting into higher-energy E$_g$ ($d_{z^2}$ and $d_{x^2-y^2}$) and lower-energy T$_{2g}$ orbitals ($d_{xy}$, $d_{xz}$, and $d_{yz}$). The Ag$^+$ cation has 10 $d$ electrons, and 6 of them occupy the T$_{2g}$ orbitals and the remaining 4 electrons fill the E$_g$ orbitals, hence all these antibonding states are fully occupied. Consequently, the band structure at the $\Gamma$ point is formed, as illustrated in Figure~\ref{band}(b). The irreps. calculated using IRVSP code~\cite{irvsp} at the $\Gamma$ and L points are also labeled in Figure~\ref{band} (b) and (d). At the $\Gamma$ point of rock-salt AgBr, the top three valance bands originate solely from Br-$p$ orbitals, forming the $\Gamma_4^-$ band representations, while the lower five valance bands originate solely from Ag-$d$ orbitals, contributing to the formation of the $\Gamma_3^+$ and $\Gamma_5^+$ band representations. Therefore, there is no $p$-$d$ orbital hybridization present at the $\Gamma$ point. At the L point, however, the highest two valence bands correspond to the L$_3^+$ irrep, and both Br-$p$ and Ag-$d$ orbitals have large contribution, as shown in Figure~\textcolor{red}{S2}.

	\subsection{Symmetry principles of $p$-$d$ orbital coupling}
	The molecular orbital (MO) theory states that for atomic orbitals to combine and form MOs, they must adhere to the principles of symmetry matching, energy proximity, and maximum overlap~\cite{PhysRev.32.186,hund1927deutung,huckel1932quantentheoretische,lennard1929electronic,zum1931quantentheoretische}.
	Among these principles, symmetry matching serves as a prerequisite, which determines whether a MO can be generated, while the other two principles influence the strength of the MOs. The principle of symmetry matching stipulates that in a crystal structure, the band representations induced by different atomic valence-electron band representations (ABRs) must coincide. Only those orbitals that possess matching band representations are eligible to hybridize and form MOs.

	The band representations at the $\Gamma$ and L points induced by Br-$p$ orbitals and Ag-$d$ orbitals in rock-salt AgBr and zinc blende AgBr are tabulated in Table~\ref{EBR}. The EBRs induced by E$_g$@4$a$, T$_{2g}$@4$a$, and T$_{1u}$@4b ABR at the $\Gamma$ point in rock-salt AgBr are $\Gamma_3^{+}$, $\Gamma_5^{+}$ and $\Gamma_4^{-}$, respectively. Due to the absence of common irreps. between Ag-related and Br-related EBRs, the coupling between corresponding Br-$p$ and Ag-$d$ orbitals is forbidden by symmetry. This is illustrated in the MO diagram at the $\Gamma$ point, where no bonding interaction occurs between the Br-$p$ and Ag-$d$ orbitals, as shown in Figure~\ref{band}. However, at the L point, the EBRs induced by E$_g$@4$a$, T$_{2g}$@4$a$, and T$_{1u}$@4b are L$_3^+$, L$_1^{+}$ $\oplus$ L$_3^+$ and L$_1^{+}$ $\oplus$ L$_3^+$, respectively. The Ag-related and Br-related EBRs share the irreps. L$_1^+$ and L$_3^+$, indicating that the coupling between Br-$p$ and Ag-$d$ orbitals is symmetry-allowed, as observed in the band structure depicted in Figure~\ref{band}. In summary, at the $\Gamma$ point, the $p$ and $d$ orbitals are symmetry-incompatible, preventing $p$-$d$ orbital coupling. In contrast, at the L point, symmetry compatibility between the $p$ and $d$ orbitals allows the formation of MOs.

	The band structure of zinc blende AgBr can also be elucidated using EBRs. As summarized in Table~\ref{EBR}, the EBRs induced by E@4d, T$_2$@4d, and T$_2$@4$a$ at the $\Gamma$ point are $\Gamma_3$, $\Gamma_4$, and $\Gamma_4$, respectively. Therefore, the T$_{2}$ orbitals of Ag and $p$-orbitals of Br exhibit strong coupling at the $\Gamma$ point, while the E$_{g}$ orbitals of Ag do not couple with Br-$p$ orbitals at this point, as illustrated in Figure~\textcolor{red}{S3}. At the L point, the EBRs induced by E@4d, T$_2$@4d, and T$_2$@4$a$ are L$_3$, L$_1$ $\oplus$ L$_3$, and L$_1$ $\oplus$ L$_3$, respectively. The common irrep. L$_1$ and L$_3$ allow the coupling between Ag-$d$ and Br-$p$ orbitals, despite the $p$-$d$ coupling at the L point is relatively weak, as shown in Figure~\textcolor{red}{S3}. Note that in addition to symmetry compatibility, other factors---such as energy alignment---plays important roles in orbital hybridization. The MO diagram at the $\Gamma$ point demonstrates that the Br-$p$ and Ag-$d$ orbitals transforming as the T$_2$ representation yield bonding and anti-bonding states, respectively, whereas the Ag-$d$ orbitals in the E representation remain non-bonding.

	The presence of the inversion center in the octahedron of rock-salt AgBr results in the ABRs of Ag-$d$ orbitals exhibiting even parity (`g') and those of Br-$p$ orbitals displaying odd parit (`u') at the $\Gamma$ point. This parity opposition leads to the EBRs induced at the $\Gamma$ point being labeled with positive (``+") and negative (``-") signs, rendering them incompatible and thereby prohibiting $p$-$d$ coupling at the $\Gamma$ point. Therefore, the determination of $p$-$d$ coupling can be made without the need for irrep. calculations, which facilitates the serach for new materials. Note the parity criterion for ABRs discussed here differs from the previous work focusing on the mixing of even and odd angular moment~\cite{PhysRevB.37.8958}. In contrast, since zinc blende AgBr lacks an inversion center, the potential for $p$-$d$ orbital coupling cannot be assessed solely based on ABRs, thereby necessitating irrep. calculations.

	\begin{figure}
		\includegraphics[width=1.0\linewidth,angle=0]{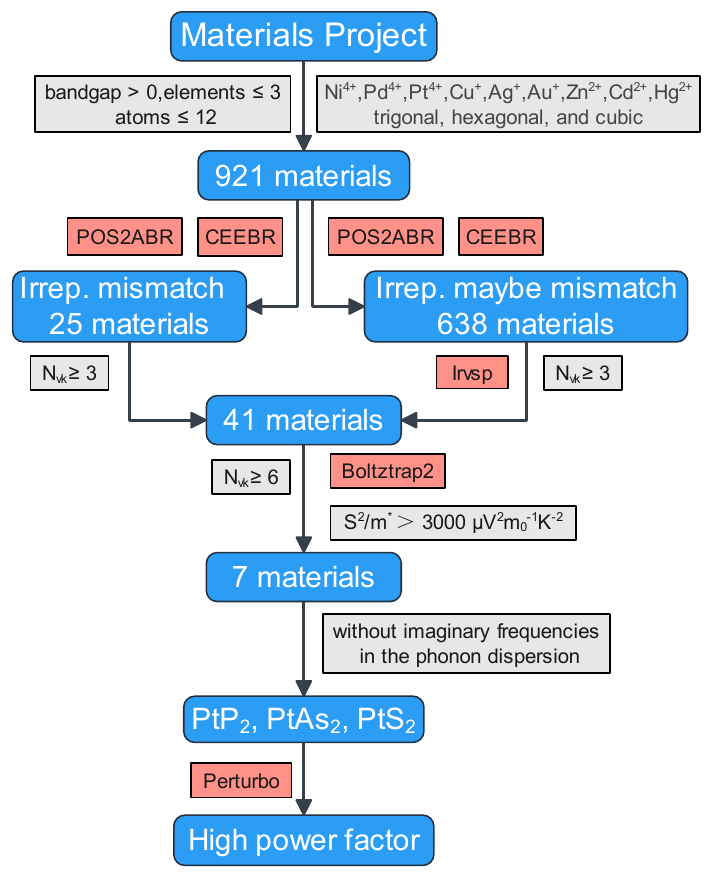}	
		\caption{Screening flow chart of materials.}
		\label{screen}
	\end{figure}

	\subsection{Electronic transport properties of rock-salt and zinc blende AgBr}
	In order to verify the role played by high $N_{\mathrm{vk}}$ in enhancing PF, we calculate $\sigma$ and $S$ of the rock-salt and zinc blende AgBr, explicitly including electron-phonon coupling, as detailed in the computational method section. As shown in Figure~\ref{AgBr}(a) and (b), the $\sigma$ of both rock-salt and zinc blende AgBr increases with rising hole concentration ($n_{\mathrm{h}}$), which is expected from the Drude model. Conversely, $S$ declines linearly as $n_{\mathrm{h}}$ increasing for both structures. For a given $n_{\mathrm{h}}$, rock-salt AgBr demonstrates higher $\sigma$ and $S$ values at the same time, leading to a higher PF, see Figure~\ref{AgBr}(c). Notably, the superior PF of rock-salt AgBr is particularly pronounced at elevated $n_{\mathrm{h}}$. The high $\sigma$ of rock-salt AgBr is attributed to its high $\mu$, which is associated with a smaller $m^*$, as depicted in Figure~\ref{AgBr}(c) and (d). The electron-phonon scattering rates near the Fermi level are very similar for both rock-salt and zinc blende AgBr, as shown in Figure~\ref{AgBr}(e), despite rock-salt AgBr has a higher $N_{\mathrm{vk}}$. Evidently, the low $m^*$ and high $S$ observed in rock-salt AgBr is due to its high $N_{\mathrm{vk}}$ ($N_{\mathrm{vk}}$=4), which contributes to a larger $m_{\mathrm{d}}^*$. Overall, these factors make the PF of rock-salt AgBr 2 $\sim$ 3 times higher than that of zinc blende AgBr, underscoring the significance of a high $N_{\mathrm{vk}}$.
		
	\begin{table*}
		\footnotesize
		\setlength{\tabcolsep}{8pt} 
		\renewcommand{\arraystretch}{1.3} 
		\centering
		\caption{Compounds with absolutely forbidden $p$-$d$ orbital coupling at the $\Gamma$ point, together with the EBRs at the $\Gamma$ point induced by their $d$ and $p$ orbitals, the $N_{\mathrm{vk}}$ of the VBM, and both the $N_{\mathrm{vk}}$ and $N_{\mathrm{vk}}$ within 50 meV below the Fermi level.}
		\begin{tabular}{ccccccc}
			\hline
			\hline
			Comp.  & Space group & EBRs($d$ orbitals) & EBRs($p$ orbitals) & $N_{\mathrm{vk}}$(VBM) & $N_{\mathrm{vk}}$(50meV) & $N_{\mathrm{v}}$(50meV) \\
			\hline
			AgCl & 221 & $\Gamma_3^+,\Gamma_5^+$ & $\Gamma_4^-$ & 3 & 3 & 6 \\
			AgBr & 221 & $\Gamma_3^+,\Gamma_5^+$ & $\Gamma_4^-$ & 3 & 3 & 6 \\
			Cu$_3$N  & 221 & $\Gamma_1^+\oplus\Gamma_3^+,\Gamma_2^+\oplus\Gamma_3^+,\Gamma_5^+, \Gamma_4^+\oplus\Gamma_5^+$ & $\Gamma_4^-$ & 4 & 4 & 12 \\ 
			Ag$_3$SI & 221 & $\Gamma_1^+\oplus\Gamma_3^+,\Gamma_2^+\oplus\Gamma_3^+,\Gamma_5^+, \Gamma_4^+\oplus\Gamma_5^+$ & $\Gamma_4^-$ & 6 & 6 & 12 \\   
			KZnF$_3$   & 221 & $\Gamma_3^+,\Gamma_5^+$ & $\Gamma_4^-,\Gamma_4^-\oplus\Gamma_5^-$ & 4 & 10 & 14 \\
			KCdF$_3$   & 221 & $\Gamma_3^+,\Gamma_5^+$ & $\Gamma_4^-,\Gamma_4^-\oplus\Gamma_5^-$ & 4 & 10 & 14 \\		
			RbCdF$_3$  & 221 & $\Gamma_3^+,\Gamma_5^+$ & $\Gamma_4^-,\Gamma_4^-\oplus\Gamma_5^-$ & 4 & 10 & 14 \\
			RbHgF$_3$  & 221 & $\Gamma_3^+,\Gamma_5^+$ & $\Gamma_4^-,\Gamma_4^-\oplus\Gamma_5^-$ & 4 & 10 & 14 \\
			CsCdF$_3$  & 221 & $\Gamma_3^+,\Gamma_5^+$ & $\Gamma_4^-,\Gamma_4^-\oplus\Gamma_5^-$ & 4 & 10 & 14 \\
			CsCdCl$_3$ & 221 & $\Gamma_3^+,\Gamma_5^+$ & $\Gamma_4^-,\Gamma_4^-\oplus\Gamma_5^-$ & 4 & 10 & 14 \\
			CsCdBr$_3$ & 221 & $\Gamma_3^+,\Gamma_5^+$ & $\Gamma_4^-,\Gamma_4^-\oplus\Gamma_5^-$ & 4 & 10 & 14 \\	
			CsHgF$_3$  & 221 & $\Gamma_3^+,\Gamma_5^+$ & $\Gamma_4^-,\Gamma_4^-\oplus\Gamma_5^-$ & 4 & 10 & 14 \\
			ZnAgF$_3$  & 221 & $\Gamma_3^+,\Gamma_5^+$ & $\Gamma_4^-,\Gamma_4^-\oplus\Gamma_5^-$ & 4 & 10 & 14 \\
			MgAgF$_3$  & 221 & $\Gamma_3^+,\Gamma_5^+$ & $\Gamma_4^-,\Gamma_4^-\oplus\Gamma_5^-$ & 6 &  9 & 18\\
			Mg$_3$ZnO$_4$ & 221 & $\Gamma_3^+,\Gamma_5^+$ & $\Gamma_4^-,\Gamma_4^-\oplus\Gamma_5^-$ & 4 & 10 & 14 \\
			Mg$_3$CdO$_4$ & 221 & $\Gamma_3^+,\Gamma_5^+$ & $\Gamma_4^-,\Gamma_4^-\oplus\Gamma_5^-$ & 4 & 10 & 14 \\
			NbAgO$_3$  & 221 & $\Gamma_3^+,\Gamma_5^+$ & $\Gamma_4^-,\Gamma_4^-\oplus\Gamma_5^-$ & 6 &  6 & 6 \\
			TaAgO$_3$  & 221 & $\Gamma_3^+,\Gamma_5^+$ & $\Gamma_4^-,\Gamma_4^-\oplus\Gamma_5^-$ & 6 &  6 & 6 \\
			CuCl & 225  & $\Gamma_3^+,\Gamma_5^+$ & $\Gamma_4^-$ &  4 &  4 &  8 \\
			CuBr & 225  & $\Gamma_3^+,\Gamma_5^+$ & $\Gamma_4^-$ &  4 &  4 &  8 \\
			AgCl & 225  & $\Gamma_3^+,\Gamma_5^+$ & $\Gamma_4^-$ &  4 &  4 &  8 \\
			AgBr & 225  & $\Gamma_3^+,\Gamma_5^+$ & $\Gamma_4^-$ &  4 &  4 &  8 \\
			AgI  & 225  & $\Gamma_3^+,\Gamma_5^+$ & $\Gamma_4^-$ &  4 &  4 &  8 \\
			ZnO  & 225  & $\Gamma_3^+,\Gamma_5^+$ & $\Gamma_4^-$ &  4 &  4 &  8 \\
			CdS  & 225  & $\Gamma_3^+,\Gamma_5^+$ & $\Gamma_4^-$ & 12 & 16 & 20 \\
			\hline
		\end{tabular}
		\label{pd1}
	\end{table*}  

	Since AgCl exhibits similar band structures as AgBr, we applied the same approach to compute electron transport properties of the rock-salt and zinc blende AgCl. The results presented in Figures~\textcolor{red}{S7} and \textcolor{red}{S8} indicate that rock-salt AgCl has a significantly larger PF than its zinc blende counterpart, further confirming the importance of high $N_{\mathrm{vk}}$ in enhancing PF. In conclusion, when comparing the zinc blende structures of AgCl and AgBr to their rock-salt counterparts, the smaller $m_{\mathrm{b}}^*$ of the rock-salt structures results in higher $\sigma$, while the high $N_{\mathrm{vk}}$ enlarges $m_{\mathrm{d}}^*$, ultimately leading to high PF. Consequently, rock-salt AgCl and AgBr exhibit higher PFs.

	\subsection{High $N_{\mathrm{vk}}$ materials screening}
	With the ingredients of achieving high PF in mind, we design a screening process based on our analysis to identify new materials with high PF from the Materials Project (\href{https://legacy.materialsproject.org/}{\textcolor{blue}{www.material project.org}})~\cite{mp}, as shown in Figure~\ref{screen}. Firstly, we restrict our search to the compounds that have anti-bonding states as their valence bands. Therefore, we focus on the cations likely to form anti-bonding states in the valence band due to $p$-$d$ coupling. These include Ni$^{4+}$, Pd$^{4+}$, Pt$^{4+}$, Cu$^{+}$, Ag$^{+}$, Au$^{+}$, Zn$^{2+}$, Cd$^{2+}$, and Hg$^{2+}$. Notably, Ni$^{4+}$, Pd$^{4+}$, and Pt$^{4+}$ possess 6 $d$-electrons that fully occupy the T$_{2g}$ orbitals within an octahedral crystal field, while Cu$^+$, Ag$^+$, Au$^+$, Zn$^{2+}$, Cd$^{2+}$, and Hg$^{2+}$ have 10 $d$-electrons, which completely occupy all the $d$ orbitals. The effective EBR at the $\Gamma$ point are then determined by combining the \webposabr program with our homemade CEEBR (Common Elements of EBR) code. For compounds displaying forbidden $p$-$d$ coupling at the $\Gamma$ point, band structure calculations are conducted to identify the value of $N_{\mathrm{vk}}$. Transport properties of all compounds with $N_{\mathrm{vk}}$ (within 50 meV below the Fermi level) greater than 6 are calculated using the BolzTraP2 code~\cite{BoltzTraP2}, without including the electron relaxation time ($\tau$). Subsequently, the PF of the compounds exhibiting high $S^{2}/m^{*}$ ($>$ 3000 $\mu{\mathrm{V}}m_0^{-1}\mathrm{K}^{-2}$) are computed using the electron-phonon coupling approach. To streamline our search, we focus exclusively on compounds that contain no more than 3 elements and fewer than 12 atoms per primitive unit cell. Given that the low symmetry space group tend to exhibit low $N_{\mathrm{vk}}$, even when the VBM is located at the boundary of the BZ, we concentrate our study on space group above the tetragonal system (space group number $>$ 142).

	\begin{table*}
		\footnotesize
		\setlength{\tabcolsep}{6pt} 
		\renewcommand{\arraystretch}{1.3} 
		\centering
		\caption{Compounds with forbidden $p$-$d$ orbital coupling at the $\Gamma$ point for the highest valence band, together with the EBRs at the $\Gamma$ point induced by their $d$ orbitals, the BRs at the $\Gamma$ point induced by their $p$ orbitals, the irrep. of the highest valence band calculated by the Irvsp, the $N_{\mathrm{vk}}$ of the VBM, and both the $N_{\mathrm{vk}}$ and $N_{\mathrm{v}}$ within 50 meV below the Fermi level.}
		\begin{tabular}{cccccccc}
			\hline
			\hline
			Comp.  & Space group & EBRs($d$ orbitals) & EBRs/BRs($p$ orbitals) & Irrep. & $N_{\mathrm{vk}}$(VBM)& $N_{\mathrm{vk}}$(50meV) & $N_{\mathrm{v}}$(50meV) \\
			\hline
			NiO$_2$  & 164 & $\Gamma_1^+,\Gamma_3^+$ & $\Gamma_1^+\oplus\Gamma_2^-,\Gamma_3^+\oplus\Gamma_3^-$ & $\Gamma_2^-$ & 6 & 12 & 12 \\
			PtS$_2$  & 164 & $\Gamma_1^+,\Gamma_3^+$ & $\Gamma_1^+\oplus\Gamma_2^-,\Gamma_3^+\oplus\Gamma_3^-$ & $\Gamma_2^-$ & 6 & 18 & 18 \\
			PtSe$_2$ & 164 & $\Gamma_1^+,\Gamma_3^+$ & $\Gamma_1^+\oplus\Gamma_2^-,\Gamma_3^+\oplus\Gamma_3^-$ & $\Gamma_2^-$ & 6 & 12 & 12 \\
			NiO$_2$  & 166 & $\Gamma_1^+,\Gamma_3^+$ & $\Gamma_1^+\oplus\Gamma_2^-,\Gamma_3^+\oplus\Gamma_3^-$ & $\Gamma_2^-$ & 6 & 12 & 12 \\
			ScAgO$_2$  & 166 & $\Gamma_1^+,\Gamma_3^+$ & $\Gamma_1^+\oplus\Gamma_2^-,\Gamma_3^+\oplus\Gamma_3^-$ & $\Gamma_3^-$ & 3 & 3 &  3 \\
			ErAgS$_2$  & 166 & $\Gamma_1^+,\Gamma_3^+$ & $\Gamma_2^-,\Gamma_3^-,\Gamma_1^+\oplus\Gamma_2^-,\Gamma_3^+\oplus\Gamma_3^-$ & $\Gamma_2^-$ & 6 & 7 & 8 \\
			TmAgS$_2$  & 166 & $\Gamma_1^+,\Gamma_3^+$ & $\Gamma_2^-,\Gamma_3^-,\Gamma_1^+\oplus\Gamma_2^-,\Gamma_3^+\oplus\Gamma_3^-$ & $\Gamma_2^-$ & 6 & 7 & 8 \\
			\multirow{2}{*}{PtP$_2$} & \multirow{2}{*}{205} & \multirow{2}{*}{$\Gamma_1^+\oplus\Gamma_4^+,\Gamma_2^+\Gamma_3^+\oplus\Gamma_4^+$} & $\Gamma_1^+\oplus\Gamma_1^-\oplus\Gamma_4^+\oplus\Gamma_4^-$, & \multirow{2}{*}{$\Gamma_4^-$} & \multirow{2}{*}{12} & \multirow{2}{*}{12} & \multirow{2}{*}{12} \\
			&&&$\Gamma_2^+\Gamma_3^+\oplus\Gamma_2^-\Gamma_3^-\oplus2\Gamma_4^+\oplus2\Gamma_4^-$ &&&&\\
			\multirow{2}{*}{PtAs$_2$} & \multirow{2}{*}{205} & \multirow{2}{*}{$\Gamma_1^+\oplus\Gamma_4^+,\Gamma_2^+\Gamma_3^+\oplus\Gamma_4^+$} & $\Gamma_1^+\oplus\Gamma_1^-\oplus\Gamma_4^+\oplus\Gamma_4^-$, & \multirow{2}{*}{$\Gamma_4^-$} & \multirow{2}{*}{12} & \multirow{2}{*}{12} & \multirow{2}{*}{12} \\
			&&&$\Gamma_2^+\Gamma_3^+\oplus\Gamma_2^-\Gamma_3^-\oplus2\Gamma_4^+\oplus2\Gamma_4^-$ &&&&\\
			Ca$_2$AuAs  & 225 & $\Gamma_3^+,\Gamma_5^+$ & $\Gamma_4^-\oplus\Gamma_5^+,\Gamma_4^-$ & $\Gamma_4^-$ & 4 & 4 & 4 \\
			Ca$_2$AuSb  & 225 & $\Gamma_3^+,\Gamma_5^+$ & $\Gamma_4^-\oplus\Gamma_5^+,\Gamma_4^-$ & $\Gamma_4^-$ & 4 & 4 & 4 \\
			Sr$_2$AuAs  & 225 & $\Gamma_3^+,\Gamma_5^+$ & $\Gamma_4^-\oplus\Gamma_5^+,\Gamma_4^-$ & $\Gamma_4^-$ & 4 & 4 & 4 \\
			Sr$_2$AuSb  & 225 & $\Gamma_3^+,\Gamma_5^+$ & $\Gamma_4^-\oplus\Gamma_5^+,\Gamma_4^-$ & $\Gamma_4^-$ & 4 & 4 & 4 \\
			Sr$_2$AuBi  & 225 & $\Gamma_3^+,\Gamma_5^+$ & $\Gamma_4^-\oplus\Gamma_5^+,\Gamma_4^-$ & $\Gamma_4^-$ & 4 & 4 & 4 \\
			Yb$_2$AuAs  & 225 & $\Gamma_3^+,\Gamma_5^+$ & $\Gamma_4^-\oplus\Gamma_5^+,\Gamma_4^-$ & $\Gamma_4^-$ & 4 & 4 & 4 \\
			Yb$_2$AuSb  & 225 & $\Gamma_3^+,\Gamma_5^+$ & $\Gamma_4^-\oplus\Gamma_5^+,\Gamma_4^-$ & $\Gamma_4^-$ & 4 & 4 & 4 \\
			\hline
		\end{tabular}
		\label{pd2}
	\end{table*} 

	Our results are summarized in Table~\ref{pd1} and Table~\ref{pd2}. For the compounds listed in Table~\ref{pd1}, the EBRs induced by cation-$d$ orbitals carry a ``+" sign, indicating spatial inversion symmetry, while those induced by anion-$p$ orbitals carry a ``-" sign, reflecting spatial inversion anti-symmetry. Since even and odd parity cannot mix, it follows that $p$-$d$ coupling is forbidden at the $\Gamma$ point for all energy bands of these compounds. The values of $N_{\mathrm{vk}}$ obtained from the search are listed in Table~\ref{pd1}, and the orbital-projected electronic structures are illustrated in Figures~\textcolor{red}{S9}-\textcolor{red}{S12}. Additionally, all compounds exhibit high $N_{\mathrm{vk}}$ values for their VBM. For example, the $N_{\mathrm{vk}}$ of Ag$_3$SI, NbAgO$_3$, and AgTaO$_3$ with space group 221 is as high as 6, and the $N_{\mathrm{vk}}$ value for CdS (with space group 225) is even as high as 12. The $N_{\mathrm{vk}}$ values of other compounds are 3 or 4.

	\begin{figure*}
		\setlength{\unitlength}{1cm}
		\includegraphics[width=1.0\linewidth,angle=0]{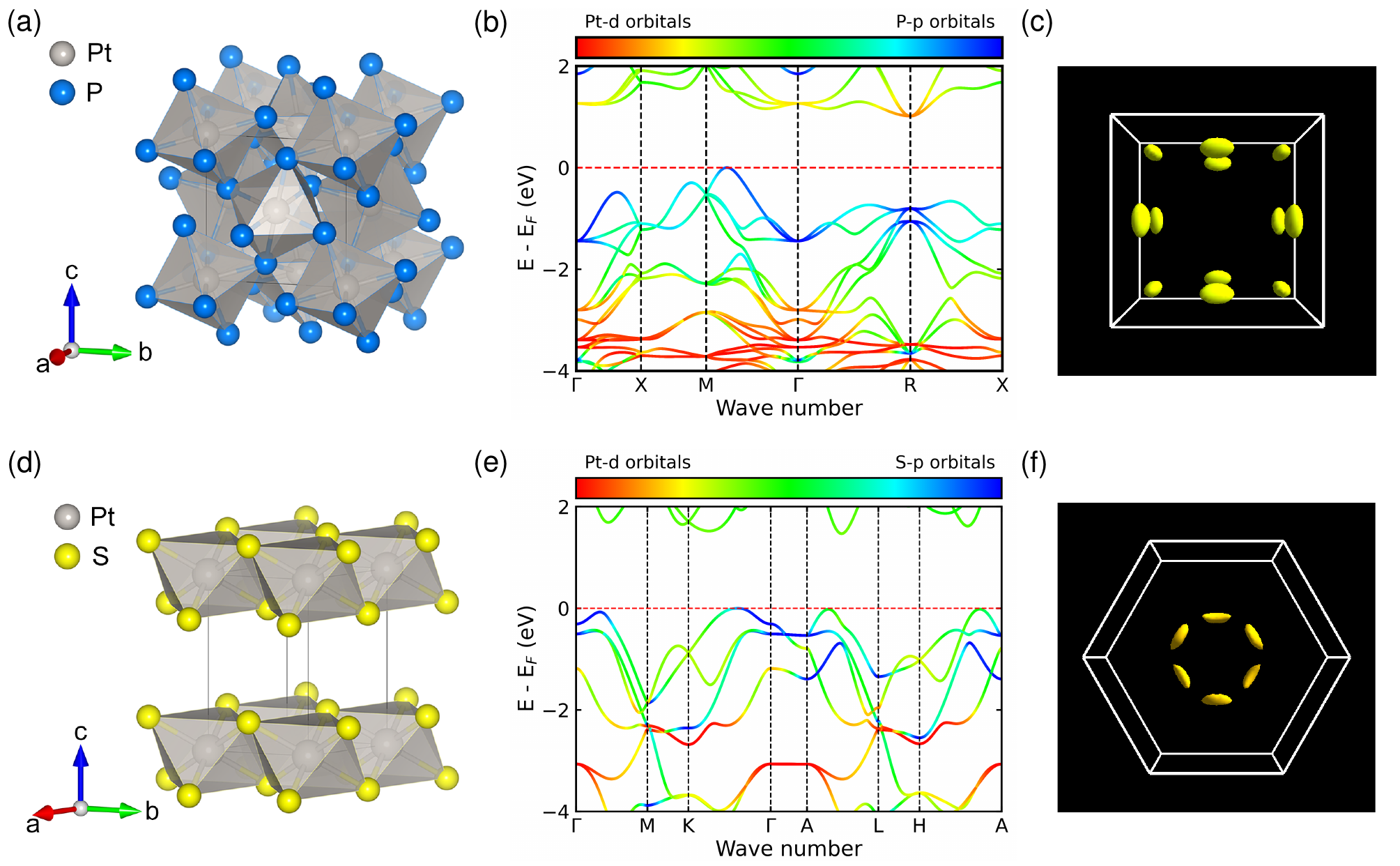}	
		\caption{(a-c) Crystal structure, orbital-projected electronic structure, and Fermi surface 10 meV below the Fermi level of PtP$_2$ with space group No. 205, respectively. (d-f) Crystal structure, orbital-projected electronic structure, and Fermi surface 10 meV below the Fermi level of PtS$_2$ with space group No. 164, respectively.}
		\label{333}
	\end{figure*}

	\begin{figure*}
		\setlength{\unitlength}{1cm}
		\includegraphics[width=1.0\linewidth,angle=0]{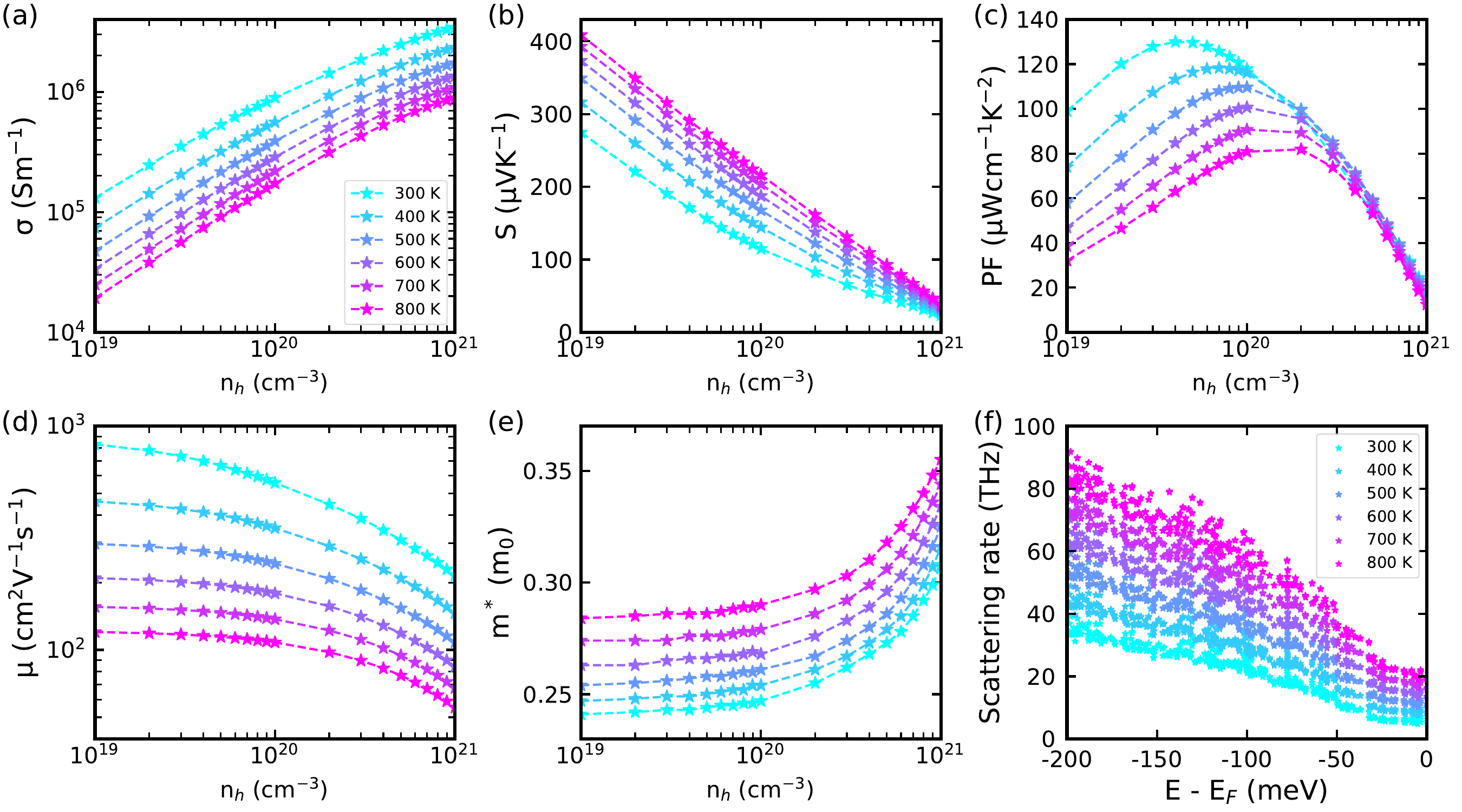}	
		\caption{The electronic transport properties, transport effective mass, and electron-phonon scattering rate of the $Ia\overline{3}$ phase of PtP$_2$. (a)-(e) The variation trends of $\sigma$, $S$, PF, $\mu$, and $m^*$ of the $Ia\overline{3}$ phase of the PtP$_2$ with $n_h$ at different $T$, respectively. (f) The variation trend of the 1/$\tau$ of the $Ia\overline{3}$ phase of the PtP$_2$ with energy at different $T$.}
		\label{PtP2}
	\end{figure*}

	The compounds listed in Table~\ref{pd2} belong to a different category, wherein $p$-$d$ orbital coupling is not completely forbidden for all energy bands at the $\Gamma$ point but is specifically prohibited at the $\Gamma$ point of the highest valence band. Take PtP$_2$ with space group 205 as an example. At the $\Gamma$ point, the EBRs induced by Pt-$d$ orbitals is $\Gamma_1^+\oplus\Gamma_4^+,\Gamma_2^+\Gamma_3^+\oplus\Gamma_4^+$, while the BRs induced by P-$p$ orbitals are $\Gamma_1^+\oplus\Gamma_1^-\oplus\Gamma_4^+\oplus\Gamma_4^-$ and $\Gamma_2^+\Gamma_3^+\oplus\Gamma_2^-\Gamma_3^-\oplus2\Gamma_4^+\oplus2\Gamma_4^-$. Due to the overlap of these two sets of EBRs or BRs in certain irreps., an analysis based solely on EBRs or BRs cannot unambiguously establish whether $p$–$d$ orbital coupling occurs. Our IRVSP calculations show that the irrep. at the $\Gamma$ point of the highest valence band is $\Gamma_4^-$, indicating that this point can only be attributed from As-$p$ orbitals, and thus $p$-$d$ orbital coupling is symmetrically prohibited. The EBRs or BRs at the $\Gamma$ point, along with valley degeneracy and corresponding orbital-projected electronic structures of these materials, are summarized in Table~\ref{pd2} and illustrated in Figures~\textcolor{red}{S13} and \textcolor{red}{S14}. Among them, for PtS$_2$ and PtSe$_2$ with space group $P\bar{3}m1$ (No. 164), as well as TmAgS$_2$ and ErAgS$_2$ with space group $R\bar{3}m$ (No. 166), the $N_{\mathrm{vk}}$ reaches a value of 6. In contrast, PtP$_2$ and PtAs$_2$, which also belong to space group 205, exhibit an even higher $N_{\mathrm{vk}}$ of 12.

	Subsequently, we calculated $S^2/m^*$ at $n_{\mathrm{h}}$ = 1 $\times$ 10$^{19}$ cm$^{-3}$ for the compounds listed in Table~\ref{pd1} and Table~\ref{pd2} with $N_{\mathrm{vk}} \geq 6$ within 50 meV below the Fermi level. The optimal values of $S^2/m^*$, along with the corresponding $S$, $m^*$, and $T$, are presented in Table~\textcolor{red}{S1}. Compounds exhibiting $S^2/m^*$ $\geq$ 3000 $\mu$V$^{2}$$m_0^{-1}$K$^{-2}$ include PtP$_2$, CdS, ErAgS$_2$, TmAgS$_2$, Ag$_3$SI, and PtS$_2$. Due to the presence of imaginary frequencies in CdS, ErAgS$_2$, TmAgS$_2$, and Ag$_3$SI at 0 K, electron-phonon coupling calculations are not feasible for these compounds. Therefore, we next calculated the PF of PtP$_2$, PtAs$_2$, and PtS$_2$ using the electron-phonon coupling method.

	\subsection{Electronic transport properties of PtP$_2$ and PtS$_2$}
	The crystal structures, orbital-projected electronic structures, and Fermi surfaces of PtP$_2$ (space group $Pa\overline{3}$, No. 205) and PtS$_2$ (space group $P\overline{3}m1$, No. 164) are shown in Figure~\ref{333}. PtP$_2$ and PtS$_2$ belong to the cubic and trigonal crystal systems, respectively, with the Pt$^{4+}$ cation coordinated in an octahedral geometry by six P$^{2-}$ or six S$^{2-}$ anions. The present of P-P bond in PtP$_2$ leads to (P$_2$)$^{4-}$ or effectively P$^{2-}$ anion. Notably, the P-Pt-P (or S-Pt-S) bond angles of both octahedra are distorted away from 180$^{\circ}$. As illustrated in Figure~\ref{333}(b) and (e), both PtP$_2$ and PtS$_2$ are indirect bandgap semiconductors. The VBM of PtP$_2$ is located between the $\Gamma$ and M points (the $\Sigma$ line), while that of PtS$_2$ lies between the $\Gamma$ and K points ($\Lambda$ line). At the VBM, PtP$_2$ exhibits 12 hole pockets, whereas PtS$_2$ has 6, as shown in Figure~\ref{333}(c) and (f), corresponding to a $N_{\mathrm{vk}}$ of 12 and 6, respectively. It is noteworthy that the $\Gamma$-K direction where the VBM of PtS$_2$ is located is an in-plane direction, which typically leads to superior properties in the in-plane orientation.

	The electronic transport properties and the electron-phonon scattering rate (1/$\tau$) of PtP$_2$ are shown in Figure~\ref{PtP2}. The $\sigma$ of PtP$_2$ increases with rising $n_{\mathrm{h}}$ and decreases with increasing $T$, while the $S$ exhibits the opposite tend, decreasing with increasing $n_{\mathrm{h}}$ and increasing with increasing $T$. These trends in $\sigma$ and $S$ are consistent with those observed in rock-salt and zinc blende AgBr, as discussed earlier. Notably, at 300 K and an $n_{\mathrm{h}}$ of $1\times10^{19}$ cm$^{-3}$, the $\sigma$ of PtP$_2$ is two orders of magnitude greater than that of rock-salt AgBr. This disparity can be attributed to the facts that the $m^*$ of PtP$_2$ is approximately 3 times smaller than that of rock-salt AgBr, and 1/$\tau$ is more than one order of magnitude lower, see Figure~\ref{AgBr}(f) and Figure~\ref{PtP2}(f). The $S$ of PtP$_2$ is approximately half that of rock-salt AgBr, as its extremely small $m^*$ limits the $S$, despite the higher $N_{\mathrm{vk}}$ value of PtP$_2$ relative to rock-salt AgBr. Due to the opposite dependence of $\sigma$ and $S$ on carrier concentration, the PF exhibits an inverted parabolic shape with respect to carrier concentration, initially increasing before decreasing, as illustrated in Figure~\ref{PtP2}(c). At 300 K, the PF of PtP$_2$ reaches a maximum value (PF$_{\mathrm{max}}$) of 130 $\mu$Wcm$^{-1}$K$^{-2}$ at an $n_{\mathrm{h}}$ of $4\times10^{19}$ cm$^{-3}$, while at 800 K, the PF$_{\mathrm{max}}$ of 82 $\mu$Wcm$^{-1}$K$^{-2}$ is achieved at $n_{\mathrm{h}}$ of $2\times10^{20}$ cm$^{-3}$. The outstanding PF of PtP$_2$ stems from its small $m^*$ and high $\mu$ (see Figure~\ref{PtP2}(d) and (e)), which lead to high $\sigma$, and its high $N_{\mathrm{vk}}$, which ensures a relatively high $S$. For the isoelectrical compound PtAs$_2$, the PF$_{\mathrm{max}}$ value at 300 K and 800 K are 127 $\mu$Wcm$^{-1}$K$^{-2}$ and 73 $\mu$Wcm$^{-1}$K$^{-2}$, respectively (see Figure~\textcolor{red}{S15}). In the case of trigonal PtS$_2$, the in-place PF$_{\mathrm{max}}$ is 82 $\mu$Wcm$^{-1}$K$^{-2}$ at 300 K and 64 $\mu$Wcm$^{-1}$K$^{-2}$ at 800K (see Figure~\textcolor{red}{S16}). These PF$_{\mathrm{max}}$ values are approximately 3 $\sim$ 5 times higher than those of high-ZT materials, such as SnSe~\cite{doi:10.1126/science.aad3749}, and even exceed that of NbFeSb (106 $\mu$Wcm$^{-1}$K$^{-2}$ at room temperature)~\cite{NbFeSbexp}, which is known to exhibit the highest PF among experimentally measured semiconductors. 
	To validate our calculations, we also compute the electrical transport properties of $p$-type half-Heusler NbFeSb using the same methodology. Our calculations indicate that NbFeSb exhibits a optimal PF of approximately 90 $\mu$Wcm$^{-1}$K$^{-2}$ at 300 K, as illustrated in Figure~\textcolor{red}{S17}. The good agreement between our calculated value, experimental date~\cite{NbFeSbexp}, and a previous theoretical value~\cite{NbFeSbtheory} supports the reliability of our computational approach.

    In summary, we propose a novel strategy of designing $p$-type semiconductors with high valley degeneracy and high power factor, based on molecular orbital theory and group theory. For compounds exhibiting strong $p$-$d$ orbital coupling, the absence of certain common symmetry elements in the elementary band representations—arising from anion-$p$ and cation-$d$ orbitals at the Brillouin zone center—gives rise to low-energy non-bonding states. Conversely, symmetry-allowed $p$-$d$ anti-bonding interactions at other $k$-points raise the energy of the valence band maximum, shifting it from the $\mathrm{\Gamma}$ to $k$-points near the zone boundaries. This shift effectively increases $N_{\mathrm{vk}}$, leading to a significant enhancement in $\mathrm{PF}$. Combining first-principles calculations with symmetry analysis based on group theory, we systematically explored 921 binary and ternary semiconductors from the Materials Project, and identified 41 compounds with $N_{\mathrm{vk}}$ $\geq$ 3 and 7 compounds with $N_{\mathrm{vk}} \geq 6$ that exhibit high $S^2/m^*$. High-accuracy DFT calculations, which fully account for electron-phonon coupling, demonstrate that PtP$_2$ ($N_{\mathrm{vk}} = 12$), PtAs$_2$ ($N_{\mathrm{vk}} = 12$), and PtS$_2$ ($N_{\mathrm{vk}} = 18$) exhibit maximum PF of 130, 127, and 82 $\mu$Wcm$^{-1}$K$^{-2}$ at 300K, respectively. These values are three to five times higher than those of well-studied TE materials and even exceed that of NbFeSb, which possesses the highest PF among known semiconductors. Our results not only elucidate the fundamental mechanisms governing orbital coupling in solids with $p$-$d$ coupling but also establish a universal framework for efficiently screening $p$-type semiconductors with high $N_{\mathrm{vk}}$ through symmetry analysis. By integrating group theory with computational materials screening, this work provides a clear pathway for the rational design of next-generation high-performance thermoelectric materials.

	\section{Computational methods}
	The transport effective mass~\cite{gibbs2017effective} is defined as $m^{*}(T,\epsilon) = \frac{e^2\tau}{\sigma(T,\epsilon)}n(T,\epsilon)$, where $e$, $\tau$, and $\epsilon$ denote the electron charge, relaxation time, and electron chemical potential, respectively. The values of $S$ and $\sigma/\tau$ were calculated using BoltzTraP2~\cite{BoltzTraP2}, based on density functional theory (DFT) calculations performed with the Vienna {\it ab initio} simulation package (VASP)~\cite{vasp1,vasp2}. The PBE functional and a plane-wave basis set with a cut-off energy of 520 eV were adopted. For all self-consistent calculations, a convergence criterion of $10^{-8}$ eV was applied. The $\Gamma$-centered $k$-point grid was designed to be sufficiently dense to ensure accurate sampling of the BZ. The crystal orbital Hamilton population (COHP) and irreducible representations (irreps.) were computed using the LOBSTER~\cite{LOBSTER} and IRVSP codes~\cite{irvsp}, respectively, both of which rely on the output of self-consistent calculations. The crystal structure and Fermi surface were visualized using VESTA~\cite{vesta} and Fermisurfer~\cite{fetmi}, respectively.
		
	Electron structures, phonon dispersions, and electron-phonon matrix elements were calculated using the Perdew-Burke-Ernzerhof (PBE)~\cite{PBE} exchange-correlation functional and Rabe-Rappe-Kaxiras-Joannopoulos (RRKJ) ultrasoft pseudopotential~\cite{RRKJ}, as implemented in the Quantum Espresso (QE) package~\cite{QE,QE2,QE3}. The plane-wave kinetic energy cutoff, BZ $k$-points grid, and the $q$-points grid are summarized in Table~\textcolor{red}{S2}. The Wannier90~\cite{wannier90} code was employed to obtain maximally localized Wannier functions. The electron transport properties, including $\sigma$, $S$, mobility ($\mu$), and electron-phonon scattering rate (1/$\tau$) were calculated by solving the Boltzmann transport equation (BTE) within the framework of the iterative approach (ITA), as implemented in the PERTURBO~\cite{Perturbo} code. Detailed information regarding the fine BZ grids utilized for both $k$-points and $q$-points is also provided in Table~\textcolor{red}{S2}.

	\section{Data availability}
	The data generated in this study is provided in the Source Data file. Source data are provided with this paper.	
	
	\section{Code availability}
	The in-house code CEEBR is available for download at https://github.com/.

	\bibliographystyle{aipnum4-2}   
	\bibliography{ref}	
		
	\section{Acknowledgement} 
	We would like to acknowledge Zhijun Wang for insightful discussions. We acknowledge the support received from the Fundamental Research Funds for the Central Universities of China (USTB) and the National Natural Science Foundation of China (Grant No. 12304115). We also appreciate the computing resources provided by the USTB MatCom at the Beijing Advanced Innovation Center for Materials Genome Engineering.

	\section{Author contribution}
    J.H. conceived and supervised the project. W.X., assisted by Z.H. and Z.X., performed materials screening and electronic transport calculations. W.X. and Z.Y. conducted symmetry analysis, and all authors participated in data interpretation. The manuscript was collaboratively written and reviewed by all authors.

	\section{Competing interests}
	The authors declare no competing interests.
	
	\section{Additional information}
	\textbf{Supplementary information} is available in the online version of the paper.
		
	\noindent \textbf{Competing interests:} The authors declare no competing financial interests.
	
\end{document}